\documentclass[journal]{IEEEtran}
\ifCLASSINFOpdf
  \usepackage[pdftex]{graphicx}
\else
\fi
\usepackage{amsmath}
\usepackage{subfigure}
\usepackage{color}
\usepackage{booktabs}
\usepackage{url}
\begin{document}
%
% paper title
% Titles are generally capitalized except for words such as a, an, and, as,
% at, but, by, for, in, nor, of, on, or, the, to and up, which are usually
% not capitalized unless they are the first or last word of the title.
% Linebreaks \\ can be used within to get better formatting as desired.
% Do not put math or special symbols in the title.
\title{s-LWSR: Super Lightweight\\ Super-Resolution Network}
%
%
% author names and IEEE memberships
% note positions of commas and nonbreaking spaces ( ~ ) LaTeX will not break
% a structure at a ~ so this keeps an author's name from being broken across
% two lines.
% use \thanks{} to gain access to the first footnote area
% a separate \thanks must be used for each paragraph as LaTeX2e's \thanks
% was not built to handle multiple paragraphs
%

\author{Biao~Li, Jiabin~Liu, Bo~Wang, Zhiquan~Qi, and Yong~Shi%\IEEEmembership% <-this % stops a space
\thanks{B. Li, Z. Qi, and Y. Shi are with the School of Economics and Management, University of Chinese Academy of Sciences, Beijing 101408, China.}
\thanks{B. Li, Jiabing Liu, Z. Qi, and Y. Shi are also with the Research Center on Fictitious Economy and Data Science, Chinese Academy of	Sciences, Beijing~100190,~China, and also with the Key Laboratory of Big Data Mining and Knowledge Management, Chinese Academy of Sciences, Beijing~100190,~China (e-mail: libiao17@mails.ucas.ac.cn; liujiabin008@126.com; qizhiquan@foxmail.com; yshi@ucas.ac.cn).}
\thanks{B. Wang is with the School of Information Technology and Management, University of International Business and Economics, Beijing 100029, China. He is currently a visiting scholar in the Department of Computer Science and Engineering, Texas A\&M University, College Station, TX 77843, USA (e-mail:wangbo@uibe.edu.cn).}
\thanks{Y. Shi is also with the College of Information Science and Technology, University of Nebraska, Omaha, NE 68182, USA.}
\thanks{Correspond author: Zhiquan Qi}
%\thanks{Biao LI was with the Department
%of Electrical and Computer Engineering, Georgia Institute of Technology, Atlanta,
%GA, 30332 USA e-mail: (see http://www.michaelshell.org/contact.html).}% <-this % stops a space
%\thanks{J. Doe and J. Doe are with Anonymous University.}% <-this % stops a space
%\thanks{Manuscript received April 19, 2005; revised August 26, 2015.}
}

\maketitle

% As a general rule, do not put math, special symbols or citations
% in the abstract or keywords.
\begin{abstract}
Deep learning (DL) architectures for super-resolution (SR) normally contain tremendous parameters, which has been regarded as the crucial advantage for obtaining satisfying performance. However, with the widespread use of mobile phones for taking and retouching photos, this character greatly hampers the deployment of DL-SR models on the mobile devices. To address this problem, in this paper, we propose a super lightweight SR network: s-LWSR. There are mainly three contributions in our work. Firstly, in order to efficiently abstract features from the low resolution image, we build an information pool to mix multi-level information from the first half part of the pipeline. Accordingly, the information pool feeds the second half part with the combination of hierarchical features from the previous layers. Secondly, we employ a compression module to further decrease the size of parameters. Intensive analysis confirms its capacity of trade-off between model complexity and accuracy. Thirdly, by revealing the specific role of activation in deep models, we remove several activation layers in our SR model to retain more information for performance improvement. Extensive experiments show that our s-LWSR, with limited parameters and operations, can achieve similar performance to other cumbersome DL-SR methods.
\end{abstract}

% Note that keywords are not normally used for peerreview papers.
\begin{IEEEkeywords}
super-resolution, lightweight, multi-level information, model compression, activation operations.
\end{IEEEkeywords}

% For peer review papers, you can put extra information on the cover
% page as needed:
% \ifCLASSOPTIONpeerreview
% \begin{center} \bfseries EDICS Category: 3-BBND \end{center}
% \fi
%
% For peerreview papers, this IEEEtran command inserts a page break and
% creates the second title. It will be ignored for other modes.
\IEEEpeerreviewmaketitle

\section{Introduction}
% The very first letter is a 2 line initial drop letter followed
% by the rest of the first word in caps.
% 
% form to use if the first word consists of a single letter:
% \IEEEPARstart{A}{demo} file is ....
% 
% form to use if you need the single drop letter followed by
% normal text (unknown if ever used by the IEEE):
% \IEEEPARstart{A}{}demo file is ....
% 
% Some journals put the first two words in caps:
% \IEEEPARstart{T}{his demo} file is ....
% 
% Here we have the typical use of a "T" for an initial drop letter
% and "HIS" in caps to complete the first word.
\IEEEPARstart{H}{ow} to recover super-resolution (SR) image from its low-resolution counterpart is a longstanding problem in image processing regime \cite{glasner2009super,yang2010image,sun2010gradient,schulter2015fast}. In this paper, we focus on the problem called single image super-resolution (SISR), which widely exists in medicine \cite{shi2013cardiac}, security and surveillance \cite{gunturk2003eigenface,zou2011very}, as well as many scenarios where high-frequency details are extremely desired.

Recently, thanks to the emergence of convolutional neural networks (CNNs), specially designed SR neural networks \cite{Chao2014Learning,Ledig2017Photo,Zhang2018Image,article2,Lim2017Enhanced,Haris_2018_CVPR} as an example-based SR method, has achieved impressive performance in terms of model accuracy. Particular, these new deep learning (DL) algorithms strive to generate satisfactory SR images with super-high peak-signal-noise-ratio (PSNR) scores than their traditional competitors \cite{Chang2004Super,glasner2009super}.

% You must have at least 2 lines in the paragraph with the drop letter
% (should never be an issue)
\begin{figure}
	\centering
	\includegraphics[scale=0.38]{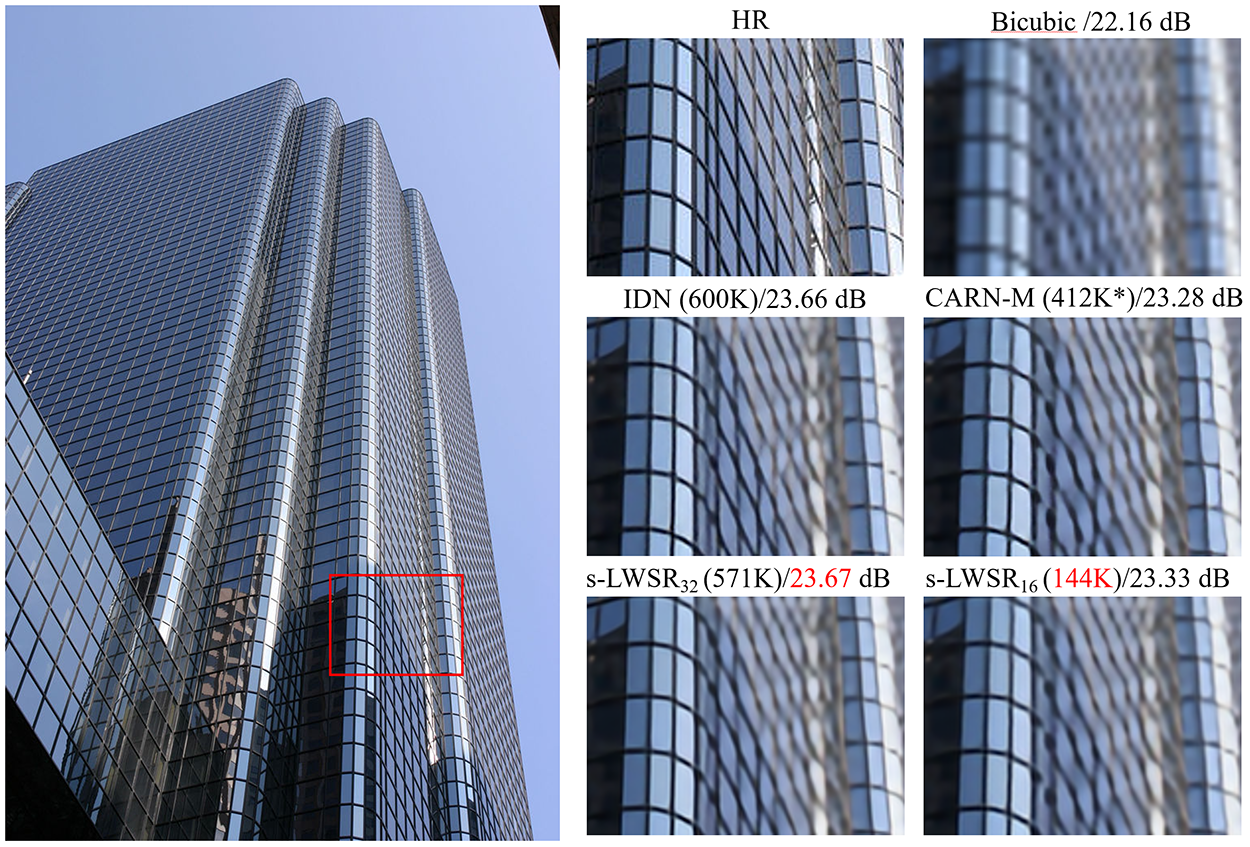}
	\caption{Visual SR results with 4X enlargement on ``img-$074$'' in benchmark dataset Urban$100$ \cite{huang2015single}. In this comparison, $s$-LWSR$_{16}$ (Ours) only uses $144k$ parameters to obtain similar performance to way larger models. Besides, if properly adding more channels in our model ($s$-LWSR$_{32}$), the final performance will surpass the others. The compared methods include: Bicubic, IDN \cite{hui2018fast}, and CARN-M \cite{article2}.}
	\label{Cont1}
\end{figure}

To the best of our knowledge, the cutting edge of SISR is CNN-based methods, normally equipped with specifically designed convolutional blocks and sophisticated mechanisms, such as global residual \cite{Kim2016Accurate}, self-attention \cite{Zhang2018Image}, Densenet \cite{huang2017densely}. In particular, the first convolutional SR network (SRCNN) is propose by Dong et al. \cite{Chao2014Learning}, based on three simple convolutional layers, where the model is trained with an interpolated low-resolution (LR) image. Although the network is not consummately design, its performance is still significantly better than almost all traditional SR algorithms. However, shallow convolutional structure constrains the model's learning ability, and the pre-processed input causes huge computation and operation cost. Hence, along with the great development of CNNs in other Computer Vision (CV) tasks, in order to leverage more information from LR inputs, SRResNet \cite{Ledig2017Photo} presents a new network by stacking $16$ residual blocks, which are learnt from ResNet \cite{he2016deep}. Later, EDSR \cite{Lim2017Enhanced} leverages $32$ modified residual blocks with $256$ channels to build an enormous SR model. Eventually, EDSR proves its super generating ability by winning the NTIRE2017 Super-Resolution Challenge \cite{timofte2017ntire} . As far as we know, RCAN \cite{Zhang2018Image} is currently the best CNN-based SR method (according to PSNR), which employs complicated residual in residual (RIR) block and self-attention mechanism.

However, as smart phones develop into regular tools for taking photos or retouch images on daily basis, CNN-based SR algorithms, which are innately designed with tremendous number of parameters, are not suitable for lightweight delivery of the model, especially as a built-in application in mobile devices. The contradiction between accuracy and efficiency raises a demanding challenge: How to deploy a CNN-based SR model on these civil-use mobile devices with a comparable performance. In other words, designing a lightweight SR network while maintaining the advanced ability in image processing becomes a rather tough yet promising computer vision task.

Generally speaking, an appropriate model architecture with well-designed hyper-parameters is needed in order to build an accurate and fast lightweight model, which is attributed to well arrangement of two principal factors therein: parameters and operations. Hence, to promote the application of SR methods on mobile devices, the essential issue will be focusing on reducing the number of parameters and operations, while keeping satisfying performance. In terms of parameters decrease, one widespread idea is to slim the network by parameters sharing among different blocks/modules. For example, the DRCN \cite{kim2016deeply} and DRRN \cite{Ying2017Image} recursively employ certain basic block with same parameters.

In addition to architecture modification, some methods attempt to reduce the operations along with the parameters through unusual convolutional layer (e.g., depth-wise separable convolution \cite{Sifre2014Rigid}), cascading structure \cite{he2016deep}, or even neural architecture search (NAS) \cite{he2018amc}. Regarding lightweight SISR, to our knowledge, CARN \cite{article2} and FALSR \cite{chu2019fast} achieve state-of-the-art results by appropriately balancing between SR restoration accuracy and model simplicity. Although these advanced compression methods have made a great progress on decreasing model size and operations, there is still a huge space for improvement. 

In this paper, we propose an adjustable super lightweight SR network called s-LWSR to promote bette balance between accurate and model size than former SR methods. The contribution of this paper is mainly threefold:
\begin{itemize}
\item Inspired by U-Net \cite{Ronneberger2015U}, we build an SR model with symmetric architecture, possessing an assistant information pool. The skip connection mechanism greatly promotes learning ability. By further combination of multi-level information from chosen layers, we build the information pool to transmit features to high-dimensional channels. Experiments show that our new architecture does well in extracting accurate information. This new information pool enforces better features transmission between the first and the second half of the model. 

\item We propose a comparatively flexible SR model compared with existing methods. Normally, the most effective factor of model size is channel numbers in intermediate layers. Here, we also modify the model size by different setting of channel numbers. Nevertheless, number change results in reduplicated model variation. Hence, by introducing a novel compression module (the inverted residual block originally borrowed from MobileNet V2 \cite{inproceedings}), the model size can be reduced by partly replacing normal residual blocks. In this way, we can control the total number of parameters within the ideal size by properly choosing the channel number and replacing specific layers with the new compression module. 
%Extensive experiments demonstrate that s-LWSR can achieve competitive results while involving limited parameters and operations.

\item According to our observation, when performing the nonlinear mapping in some activation layers (e.g., ReLU), useful information is likely to be partly discarded. As a result, we remove some activation operations to retain object details in our lightweight model. Experiments prove that this minor modification improves the performance of our lightweight SR model.
\end{itemize}

% needed in second column of first page if using \IEEEpubid
%\IEEEpubidadjcol

\section{Related Work}
With the development of deep learning, a bunch of achievements on SR has been obtained \cite{Chao2014Learning,Zhang2018Image,Ledig2017Photo,article2,chu2019fast,Lai2017Deep,Lim2017Enhanced,Haris_2018_CVPR,he2016deep,Kim2016Accurate,inproceedings,Ying2017Image,shi2019unsupervised}. There are many detailed reviews about SR development in these papers. Based on these surveys, we firstly present a brief introduction about DL-SR algorithms. Additionally, literature study addresses model compression will be given in Section \ref{MC}.

\subsection{Deep Single Image Super-Resolution (SISR)}
The first deep SISR model that surpasses almost all former traditional methods is SRCNN \cite{Chao2014Learning}. In this end-to-end network, three convolutional layers are employed to produce HR images from their interpolated LR counterparts. Then, Dong et al. push the envelope further by introducing a new architecture FSRCNN \cite{10.1007/978-3-319-46475-6_25}. The model replaces the pre-upsampling layer at the beginning of the network with a learnable scale-up layer at the end of the network. Because of training with smaller patches in most intermediate layers, the computational and operational costs greatly drop.

Subsequently, more sophisticated and powerful approaches have been proposed. For instance, by using $20$ convolutional layers and a global residual, VDSR \cite{Kim2016Accurate} obtains a shocking result that satisfies various applications. Meanwhile, DRCN \cite{kim2016deeply} proposes a deeper recursive architecture with fewer parameters. In particular, several identical layers are stacked recursively in DRCN. At the same time, recursive-supervision and skip-connection are applied to ease the problem of mis-convergence. 

Besides, benefiting from ResNet \cite{he2016deep}, SRResNet \cite{Ledig2017Photo} improves the model efficiency by stacking several residual blocks. Based on SRResNet, Lim et al. propose the EDSR \cite{Lim2017Enhanced}, which removes the batch normalization \cite{ioffe2015batch} module and expends the width of channels. However, there are still more than $40$ million parameters in this model. Recently, a very deep residual network RCAN is proposed \cite{Zhang2018Image}, which introduces a novel local block and the channel attention mechanism. As described in the paper, the attention mechanism further facilitates learning in high-frequency information. Although these methods receive the state-of-the-art results on PSNR, too many parameters ($\sim30-40$ million parameters) make them hard to run on common CPU-based computers, not to mention any mobile devices/phones.

On the other hand, although most SR algorithms persist in obtaining SOTA results in pixel level, it is still controversial that high PSNR or SSIM guarantees satisfying and realistic feeling in visual. Based on this consideration, some former researches focus on how to generate perceptual satisfying images. For example, SRGAN \cite{Ledig2017Photo} leverage the generative adversarial networks (GANs) \cite{goodfellow2014generative} with SRResNet as the generator to produce photo-realistic images. Similar to SRGAN, EnhanceNet \cite{sajjadi2017enhancenet} produces automated texture synthesis in a GANs framework. Although GAN-based SR models work well on perceptual generation, they act poorly on PSNR or SSIM accuracy.

In this paper, we mainly focus on how to obtain more accurate SR images in pixel level. However, our perspective is to properly balance between the pixel level fidelity and the model size.

\subsection{Model Compression}\label{MC}
Recently, how to make deep models be capable in running on mobile devices has received much attention. In this section, we provide a brief survey on compression methods, especially in SR relevant models. Firstly, most compression methods try to compress the model by modifying the network structure, such as \cite{zhang2018shufflenet,changpinyo2017power,article,Sandler2018Inverted}. In MobileNetV1 \cite{article}, it reduces the number of parameters through utilizing depth-wise separable convolutions \cite{Sifre2014Rigid}. Since convolution operation are separated into two steps, the total number of parameters is reduced in a large margin, accompanying with the learning ability decline. In order to maintain the accuracy as reducing the model size, MobileNetV2 \cite{Sandler2018Inverted} proposes a novel layer module: the inverted residual with linear bottleneck. A scale factor is introduced to add more channels into the compression module. As a result, we can obtain better performance by reducing the compression level. In addition, a new compression pattern: neural architecture search (NAS) \cite{zoph2018learning}, which searches architecture by genetic algorithms, reinforcement learning, and Bayesian optimization, has received much attention. In this paper, we employ a similar mechanism as MobileNetV2 to build an efficient lightweight model. 

For SR compression, Kim et al. introduce the recursive layers to share parameters in different blocks. They propose a very deep convolutional network (DRCN) \cite{kim2016deeply} consisting of $16$ identical intermediate layers. In this way, the number of parameters can be controlled when more layers are added. Similar to DRCN, DRRN \cite{Ying2017Image} utilizes both global and local residual learning to further optimize the method. Using these recursive blocks, DRRN with $52$ recursive layers surpasses former methods in performance. Recently, Ahn et al. design an efficient and lightweight model called CARN \cite{article2}. Their compression strategies include the residual-E (similar to MobileNetV1 \cite{article}) and the recursive layers in the cascading framework. Finally, the CARN-M achieves comparable accuracy to other CNN-based SR methods, with fewer parameters and operations than CARN. Besides, the NAS strategy (like \cite{zoph2018learning}) is proposed in FALSR \cite{chu2019fast}. Unsurprisingly, its result is comparable with CARN or CARN-M with appropriate model size. However, the generated architecture is extremely complex and hard to explain. Besides, Ma et al. make efforts to use binary weights and operations, compared with general 16-bit or 32-bit float operations, to address the over-parametrization in \cite{ma2018efficient}.

Though these lightweight SR models have achieved great success, there is still huge improvement space in how to obtain a better balanced and more flexible SR model. This is the start point of our research.

\section{Methodology}
\begin{figure*}[ht]
	\centering
	\includegraphics[scale=0.4]{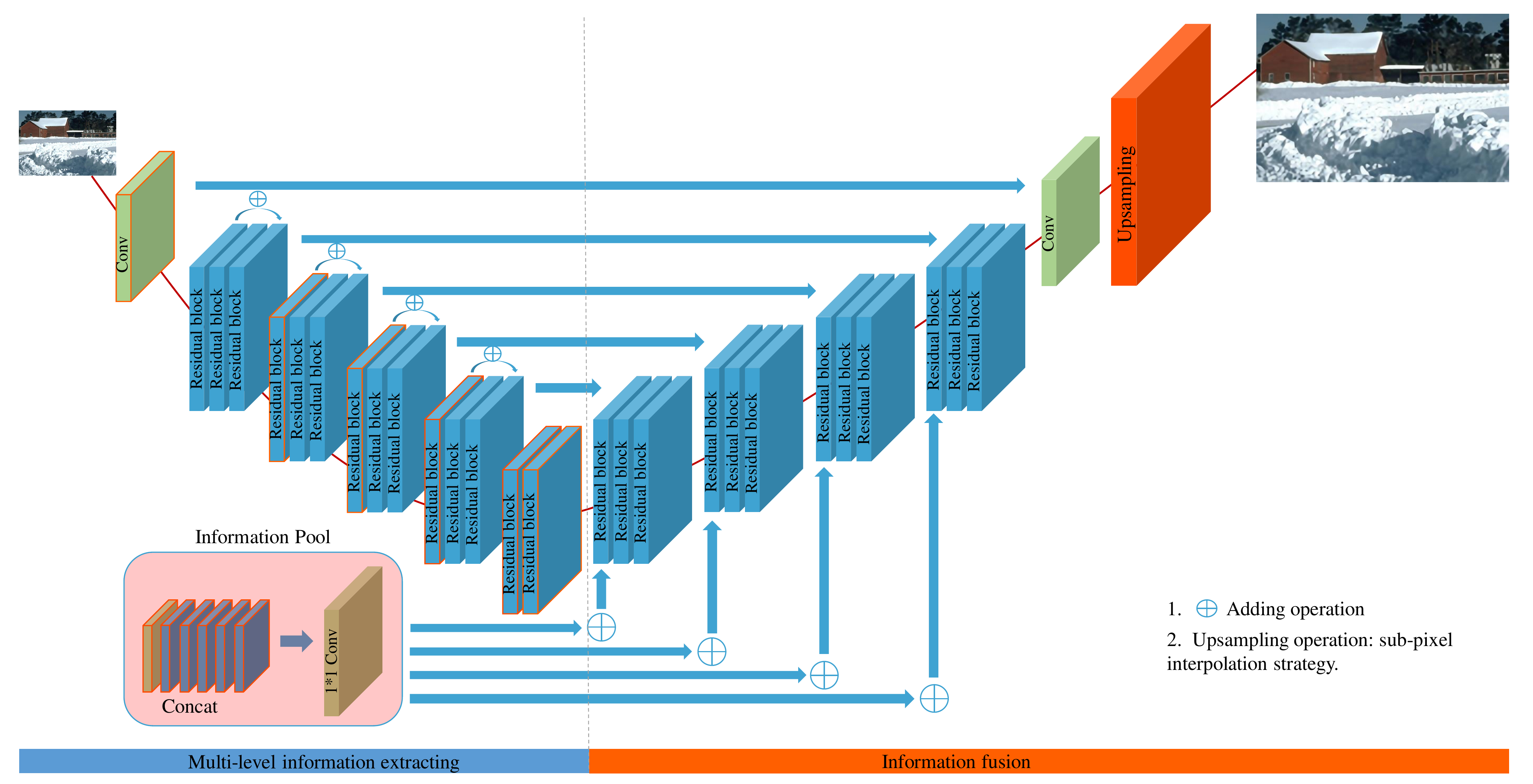}
	\caption{The architecture of s-LWSR. The blue one is the basic residual block with all chosen blocks for information pool marked in red. Convolutional layers appear in green color. The information pool and the path of information are also marked with arrow lines.}
	\label{Total structure}
\end{figure*}

In this section, we present the technical details of s-LWSR, which consists of five parts: basic residual blocks, symmetric connection frame, information pool, model compression, and activation removal mechanism. The first part, residual blocks, is the fundamental unit used to sufficiently extract information from the LR image (i.e., $I^{LR}$). The second and third parts work as the backbone of the network, functioning as the fusion of multi-level information among intermediate layers. In the fourth part, we further introduce a compression module to decrease the number of parameters and operations, so that the model size can be controlled within an ideal range. In the last part, selected activation layers are removed from the pipeline to retain more information in inner layers. The architecture of our s-LWSR is shown in Fig. \ref{Total structure}.

\subsection{Basic Residual Block}
\begin{figure}[ht]
	\centering
	\includegraphics[scale=0.45]{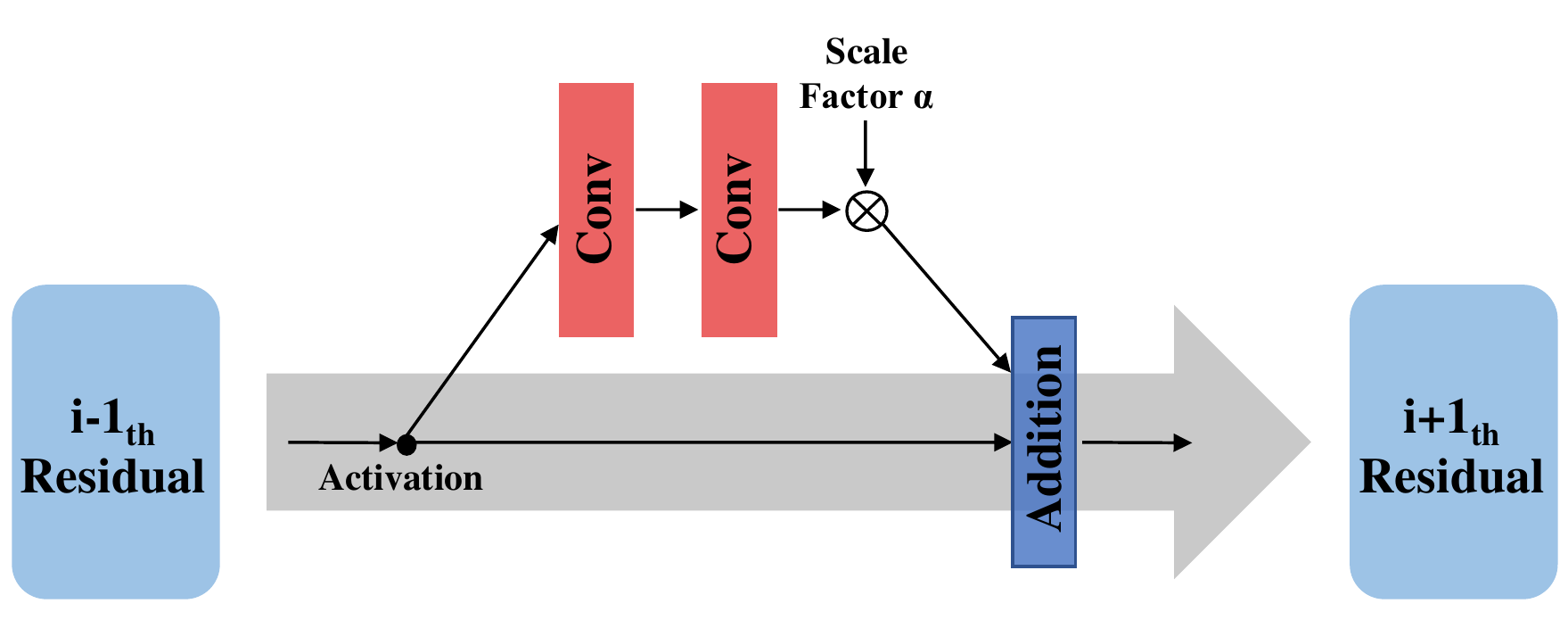}
	\caption{The proposed residual block in s-LWSR. There are two separate information flows. A scale factor is used to control the magnitude of the information introduced in the branch. Features from two flows are element-wise added as the input of the next block.}
	\label{Residual}
\end{figure}

We firstly introduce the basic cell of s-LWSR: the residual block ($\mathcal{R}$)\cite{he2016deep}, which plays the fundamental role in our model. It leads to excellent extracting ability as learning from the LR inputs ($I^{LR}$). The $i_{th}$ cell is defined as:
\begin{equation}
\begin{aligned}
&R^{i} = \alpha \cdot Conv(Conv(\mathcal{F}(R^{i-1}))) \cdot  + \mathcal{F}(R^{i-1}),  %\nonumber
\end{aligned}
\end{equation}
where $R^{i}$ refers to final output of the $i_{th}$ residual block. As shown in Fig. \ref{Residual}, the starting activation operation $(\mathcal{F})$ is utilized to process initial input to all following operations. In the branch part, two convolutional layers are cascaded like other residual setting. A scale factor: $\alpha$ is introduced to control the effect of residual branch. Both of them are used to extract useful information and increase dimensions. Inspired by the EDSR, we remove all batch normalization layers from the original residual block to enhance final performance, as well as reducing the redundant operations.

\subsection{Symmetric Connection Frame}
Inspired by U-Net \cite{Ronneberger2015U}, we propose a novel symmetric architecture which is depicted in Fig. \ref{Total structure}. Like most SR models, the whole process of s-LWSR contains three sub-procedures: original feature extraction, detailed information learning, and SR image restoration. The RBG inputs ($I_{LR}$) are firstly operated by original feature extraction part. Then, pre-processed layers go through a series of well-designed blocks which are used to act accurate information. Finally, SR images ($I_{SR}$) are generated from the last outputs containing abundant features by the SR image restoration block, where HR images ($I_{HR}$) supervise the quality of generations. 

In s-LWSR, experiments prove a trade-off between accuracy and model size: the more channels involved, the better performance achieved. In order to flexibly adjust the model size, we set the channel number of all residual blocks, n-feats ($\beta$), as the primary factor of model size.  In Fig. \ref{Total structure}, the channel number is chosen from $[16, 32, 64, 128]$. With the increasing of $\beta$, the model size enlarges diploid. More experimental details about models with different channels are shown in Section \ref{details}.

As shown in Fig. \ref{Total structure}, a sequence of basic residual blocks consecutively connected, aiming at learning the feature map between $I^{LR}$ and $I^{SR}$. Similar to U-Net, our model equips the skip-connection between corresponding structure channels, and the entire mid-procedure is separated into nine bunches of local blocks ($LB_{s}$). Separated by function, the first five $LB_{s}$ serve as the multi-level information extractor for the information pool, and rest blocks are information fusion part. Inspired by RDN \cite{zhang2018residual}, we further introduce local residual learning (LRL) to fuse features from different dimension. Given any $LB^{i}_{s}$ in the information extractor, the information propagating process runs as follows: 
\begin{equation}
\centering
    \begin{aligned}
    %\left\{
    \begin{array}{lr}
    LB^{i}_{output} = LB_{R3}^{i}(LB_{R1}^{i}(LB^{i-1}_{output}) + LB_{R2}^{i}(LB_{R1}^{i})). &\\%    i \in [2,3,4,5].& %\nonumber
    \end{array}
    %\right.
    \end{aligned}
\end{equation}
Benefit from the skip-connection and LRL, the $I^{LR}$ can be sufficiently processed in local spatial architecture with multi-level information. 

For the latter half of $LB_{s}$, the sum of features from the information pool and skip connection of former layers form their input. To coordinate the proportion, we set $0.5$ as weight for either source. As a result, s-LWSR is not only fully extracting multi-level information from information pool, but also fully utilizes specific features of its corresponding former layers.

\subsection{Information Pool}
For combining detailed multi-layer information, specific layers in the former five $LB_{s}$ are chosen as sources of the information pool. As shown in Fig. \ref{Total structure}, we mark these layers with red border. All chosen layers are firstly concatenated, and then followed with a $1 \times 1$ convolutional layer which is used to reduce these five times concatenated layers to original input numbers. Finally, the output of information pool contains the same number of channels as other residual blocks. To be specified, equal layers is the basic processing for adding operations at any point of the network. In general, the whole process of information pool can be described as:
\begin{equation}
    \begin{aligned}
    IP_{output} = {Conv}^{*}(Cat[conv_{1}, R^{1}_2, R^{1}_3, R^{1}_4, R^{1}_5, R^{2}_5]), %\nonumber
    \end{aligned}
\end{equation}
where ${Conv}^{*}$ denotes the $1 \times 1$ convolution, and $R^i_j$ represents the $i_{th}$ residual block in the $j_{th}$ block bunch.

In fact, a similar structure has been introduced in DRCN \cite{inproceedings}, where all predictions from different layers are weighted combined in the last layer. The intention therein is to train the network in a supervised way. The output of inner blocks is summed with an extra weight factor $w$. Then, the output $I^{SR}$ is determined by the learning ability of middle blocks, and parameter sharing is employed in all the learning blocks of DRCN for reducing the number of parameters. 

Hence, although the information pool utilizes the similar structure as DRCN, the underlying mechanism is fairly different. Instead of adding every generation in the halfway blocks, some specified dimensional layers chosen by experiments are concatenated in the information pool, which considerable alleviate the over-fitting problem. We choose the channel concatenation because it can maintain more multi-level information within the channels, whereas the channel addition operation will change the value in the tensor. Totally, the information pool introduced here is distinct from the existing information fusion strategies.

\subsection{Model Compression of s-LWSR}\label{com}
\begin{figure}[ht]
	\centering
	\includegraphics[scale=0.8]{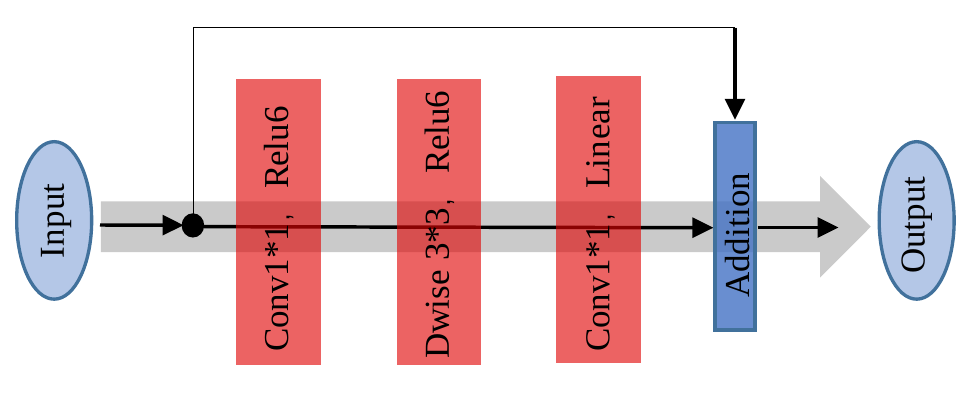}
	\caption{Illustration of the inverted residual module which is introduced in MobileNetV2. Three convolutional layers and a residual connection are involved. To improve the learning ability of the module, the channels of middle layers are increased by $1 \times 1$ convolution. Compared with MobileNetV1 \cite{article}, the number of inner layers affects the performance and total parameters.}
	\label{MobileV2}
\end{figure}

In deep learning architectures, the function of how to count parameters in each convolutional layer is like:
\begin{equation}
    Para_{sum} = F_{kernel} \times F_{kernel} \times C_{input} \times C_{output} + C_{output},
\end{equation}
where $F_{kernel}$ is the kernel size and $C$ is the channel number. In particular, when the channel number reduces by half, both of $C_{input}$ and $C_{output}$ decrease by half, which further cause that the total number of parameters approximately decreases to one quarter of its full size. On the other hand, in order to endow the model size with the flexibility, we further compress s-LWSR with a novel module: The inverted residual with linear bottleneck, which is originally introduced in MobileNetV2 \cite{Sandler2018Inverted}. This paper demonstrates that this compression module improves the performance in a large margin, compared with the depth-wise separable convolution in MobileNetV1 \cite{article}. Details of the module are illustrated in Fig. \ref{MobileV2}. In our model, some basic blocks are changed with this new module to progressively reduce the model size to the ideal range.

\subsection{Activation Removal}
To maintain more information when performing model compression, we modify s-LWSR with activation layers removal mechanism. Unlike high-level CV tasks, such as object detection YOLOV3 \cite{Redmon2018YOLOv3} and semantic segmentation \cite{chen2018encoder}, the SR task requires to recover information from the $I^{LR}$ as much as possible. Thus, maintaining the comprehensive details flow from the original input is essential to the following processing on features. However, the activation operations, e.g., ReLU, alter the details in feature map in order to realize the non-linearity, which may undermine the fidelity of useful information \cite{wang2018esrgan}. The learning ability of SR models inevitably suffers from the model compression module in a certain degree. Hence, removing some activation layers could be a proper strategy to offset the information loss brought by the model compression, and retain important feature information. Meanwhile, this operation can further reduce the computational complexity. However, it is still an open question that how many activation layers should be removed, and we strive for making it clear through looking at the influence arisen by this removal with our multi-level ablation analysis in Section \ref{MA}.

\section{Experiments}
\subsection{Implementation and Training Details}\label{details}
To fair compare our approach with other DL-SR methods, we conduct the training process on a widely used dataset, DIV2K \cite{agustsson2017ntire}, which contains $800$ LR-HR image pairs. Then, we investigate the performance of different algorithms upon four standard datasets: $Set5$ \cite{bevilacqua2012low}, $Set14$ \cite{zeyde2010single}, $B100$ \cite{martin2001database}, and $Urban100$ \cite{huang2015single}. Besides, the generated SR images are transformed into $YCbCr$ space, where we compute the corresponding PSNR and SSIM \cite{wang2004image} on the $Y$ channel.

In detail, the data augmentation is firstly adopted to the training data to improve the generalization ability. During the training process, our algorithm extracts features with $48 \times 48$ patches from the $I^{LR}$, and the objective is optimized with the ADAM ($\beta_{1}\!=\!0.9, \beta_{2}\!=\! 0.999$) \cite{kingma2014adam}. Besides, most filters in the pipeline are designed with the same size $3 \times 3$, except some $1 \times 1$ layers for channels reduction, and the learning rate is set as $1 \times 10^{-4}$, halved every $200$ epochs. We implement s-LWSR on Pytorch with a Titan Xp GPU. Our code is availabe on \url{https://github.com/Sudo-Biao/s-LWSR}.
\begin{figure*}[ht]
	\centering
	\includegraphics[scale=0.32]{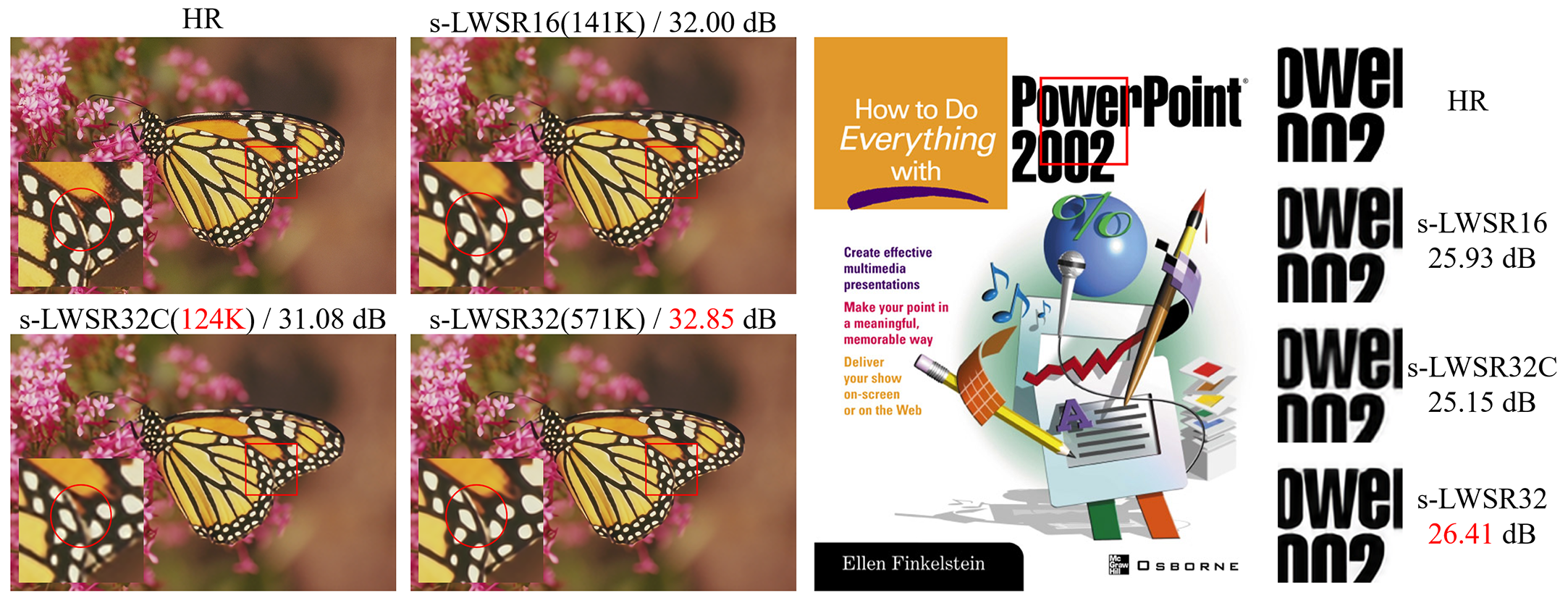}
	\caption{The comparison of s-LWSR with different model settings. The test images are from $Set14$ \cite{zeyde2010single}. In the comparison, we choose three models: s-LWSR$_{16}$,  s-LWSR$_{32}$, and compressed s-LWSR$_{32}$ (s-LWSR32C). The final performance suggests that the number of channels is a crucial factor to SR results, while the use of compression modules significantly decreases the learning ability of the model.}
	\label{cont3}
\end{figure*}

\subsection{Model Analysis}\label{MA}
Most DL-SR algorithms can be separated into three parts: feature extraction, feature learning, and up-sampling. For the first part in our method, a $conv(3,n\!-\!feats)$ layer is implemented to primarily learn the comprehensive features, which are the inputs of the next layer, the information pool, and the global residual unit. In order to maintain more details from the input and deliver them to the following operation layers, we only use one $conv(3,n\!-\!feats)$ to achieve channel number change. We will explicitly illustrate the feature learning part in Section \ref{as}. For the up-sampling part, we adopt the sub-pixel shuffling strategy, which is commonly used by other outstanding DL-SR methods.

As we mentioned in Section \ref{com}, the channel number is a crucial factor with a great effect on the model size and accuracy performance. In our experiment, we firstly use $16$ channels for the simplicity of the desirable lightweight model. Then, channels in all modules are $2 \times$ added for better learning ability, like $32$ and $64$. For the flexible parameter modification, we utilize the inverted residual module and remove some activation layers. Further analysis on the trade-off between the number of parameters and the model accuracy is provided in Fig. \ref{cont3}.

{\bfseries Channel Size.}
To demonstrate the learning ability of our model, we build several models with different $n\!-\!feats$: $16\times$, $32\times$, and $64\times$. The total number of parameters ranges from $140K$ to $2277K$. Referring to the $4\times$ SR task, s-LWSR$_{16}$ leverages an extremely small network to learn the feature map between $I^{LR}$ and $I^{SR}$, and the final result is comparable to some DL-SR methods with several times larger in parameters as shown in Table \ref{tab1}. Hence, s-LWSR$_{16}$ is the specific model that perfectly solves the mobile device implementation issue aforementioned. More visual detail comparisons are illustrated in Fig. \ref{Cont1}.

The numerical comparison can be found in Table \ref{tab2}. Experiments clearly demonstrate that the PSNR value can be significantly improved with additional parameters. However, the comparison with former leading methods proves that our model can achieve similar performance with considerable fewer parameters. In detail, we first compare our smallest model (s-LWSR$_{16}$), which is equipped with a deeper but thinner network, with other outstanding methods. Our method contains fewer parameters and operations than that of LapSRN, VDSR, and DRCN, while receiving even higher PSNR values in the final results. To be specific, for $4\times$ SR task on $Set5$, s-LWSR$_{16}$ achieves $31.62$ dB, which is respectively $0.08$ dB, $0.27$ dB, and $0.09$ dB higher than LapSRN, VDSR, and DRCN. On the other hand, the model parameter size of s-LWSR$_{16}$ is respectively $17.7\%$, $21.7\%$, and $8.1\%$ of those state-of-the-art DL-SR methods. Meanwhile, the decrease of operations is even much greater, which are $5.6\%$ of LapSRN, $1.4\%$ of VDSR, and $0.085\%$ of DRCN, respectively.

Besides, if we double the $n\!-\!feats$ to generate a bigger model: s-LWSR$_{32}$, it achieves the best performance of all SOTA DL-SR methods that with $<1000K$ parameters on datasets: $Set5$, $B100$, and $Urban100$. Compared with s-LWSR$_{16}$, s-LWSR$_{32}$ is four times larger, which leads to $0.42$ dB improvement in the final result on $Set5$. Besides, compared with former leading lightweight methods: CARN-M and IDN \cite{hui2018fast}, our $32 n\!-\!feats$ model performs better with $0.12$ dB and $0.22$ dB higher in PSNR for $4\times$ SR task on $Set5$ respectively. However, there is no data to compete with FALSR-A due to the lack of available code in public. Hence, we follow the allegation in the paper that their results are comparable to CARN-M. In particular, CARN-M proposes a single model for $2\times$, $3\times$, and $4\times$ SR images at the same time. However, when calculating the parameter and multi-adds, they divide the total parameters number by $3$. Our s-LWSR$_{32}$ contains less than half of the number of parameters and multi-adds in the entire CARN-M model, while obtaining better performance. In general, the generations of s-LWSR$_{32}$ verify the promising learning ability of the proposed set of mechanisms in our SR structure. To further study the relationship between the number of channels and the performance in our method, we increase the channel numbers to $64$, that is, s-LWSR$_{64}$. We conduct additional experiments to affirm the expected capacity of the proposed unit. The final results are displayed in Table \ref{tab1}.

{\bfseries Further Compression.}
The former comparison of s-LWSR with different $n\!-\!feats$ verifies the effectiveness and efficiency of our network. When designing a model for a practical SR problem, the number of $n\!-\!feats$ is determined by the computation resource. In addition, parameters decrease in three quarters when the $n\!-\!feat$ is halved down. There is still a huge space for the better trade-off between the number of parameters and the final accuracy. To address the issue, we introduce the inverted residual blocks derived from the MobileNetV2 in our model. When the basic residual blocks are replaced by this compression unit, the number of parameters is further reduced in a relatively small degree compared with channel changing. Taking s-LWSR$_{32}$ for an example, the total number of parameters reduces from 571K to 124K when all layers are replaced with this new module, which is a similar size as that of s-LWSR$_{16}$. We show the setting details in Table. \ref{tab2}. On the other hand, experiments also demonstrate that the reverse residual block is less capable in extracting features than the original residual block. For example, the PSNR value of the entire compressed s-LWSR$_{32}$ is $0.4$ dB less than that of s-LWSR$_{16}$ on the condition of similar model size. The comparison is shown in Fig. \ref{cont3}. As a result, the number of compressed blocks involved in the model should be elaborately determined to balance the model size and performance.

\begin{table}[!htbp]
	\begin{center}
		\small
		\caption{The comparison of the original s-LWSR and two derivatives transformed in the depth or the width. The changes of parameters and PSRN are illustrated.}
		\label{tab2}
		\begin{tabular}{l|lll}
		\hline
			Options       & s-LWSR(base line) & Depth    & Width  \\
			\hline
			Basic blocks  & 26                & 6        & 26     \\
			n-feats       & 32                & 32       & 16     \\
			Loss function & L1                & L1       & L1     \\
			Parameters    & $571K$            & $308K$   & $144K$ \\
			PSNR(+)       & $32.15$           & $31.93$  & $31.78$\\
		\hline
		\end{tabular}
	\end{center}
\end{table}

\begin{table*}
	\small	
	\centering
	\caption{The comparison of s-LWSR and other state-of-the-art methods: SRCNN \cite{Chao2014Learning}, FSRCNN \cite{10.1007/978-3-319-46475-6_25}, CARN\cite{article2}, VDSR \cite{Kim2016Accurate}, MemNet \cite{tai2017memnet}, IDN \cite{hui2018fast}, LapSRN \cite{Lai2017Deep}, DRCN \cite{kim2016deeply}, DBPN \cite{Haris_2018_CVPR}, and EDSR \cite{Lim2017Enhanced} on $4\times$ enlargement task. The PSNR and SSIM are compared according to the final results. $s-LWSR+$ denote self-ensemble versions of s-LWSR.}
	\label{tab1}   	
	\begin{tabular}{cccccccccccc}
		\toprule[2pt]
		&&&& \multicolumn{2}{c}{Set5} & \multicolumn{2}{c}{Set14} & \multicolumn{2}{c}{B100} & \multicolumn{2}{c}{Urban100}\\
		Algorithm &	Scale & Params (K) & Multi-Adds (G) & PSNR & SSIM & PSNR & SSIM & PSNR & SSIM & PSNR & SSIM \\
		\midrule[1.5pt]
		Bicubic  & $4$ & -       &       -   & $28.42$ & $0.810$ & $26.10$ & $0.704$ & $25.96$ & $0.669$ & $23.15$ & $0.659$ \\
		FSRCNN   & $4$ & $12$   & $4.6$    &$30.71$ & $0.866$ & $27.59$ & $0.753$ & $26.98$ & $0.715$ & $24.62$ & $0.728$ \\
		SRCNN    & $4$ & $57$   & $52.7$   &$30.48$ & $0.863$ & $27.49$ & $0.750$ & $26.90$ & $0.710$ & $24.52$ & $0.722$ \\
		$s$-LWSR$_{16}$(Ours)  & $4$ & $144$   & $8.3$   &$31.62$ & $0.886$ & $27.92$ & $0.770$ & $27.35$ & $0.729$ & $25.36$ & $0.762$ \\
		$s$-LWSR$_{16}+$(Ours) & $4$ & $144$   & $8.3$   &$31.78$ & $0.889$ & $28.00$ & $0.772$ & $27.40$ & $0.730$ & $25.45$ & $0.765$ \\
		CARN-M   & $4$ & $412^{*}$  & $18.3$   &$31.92$ & $0.890$ & $28.42$ & $0.776$ & $27.44$ & $0.730$ & $25.63$ & $0.769$ \\ 
		\midrule[0.5pt]        
		$s$-LWSR$_{32}$(Ours)  & $4$ & $571$   & $32.9$   &$32.04$ & $0.893$ & $28.15$ & $0.776$ & $27.52$ & $0.734$ & $25.87$ & $0.779$ \\
		$s$-LWSR$_{32}+$(Ours) & $4$ & $571$   & $32.9$   &$32.15$ & $0.894$ & $28.24$ & $0.778$ & $27.58$ & $0.736$ & $26.00$ & $0.782$ \\
		IDN      & $4$ & $600$  & $34.5$   &$31.82$ & $0.890$ & $28.25$ & $0.773$ & $27.41$ & $0.730$ & $25.41$ & $0.763$ \\       
		VDSR     & $4$ & $665$  & $612.6$  &$31.35$ & $0.884$ & $28.01$ & $0.767$ & $27.29$ & $0.725$ & $25.18$ & $0.752$ \\
		MemNet   & $4$ & $677$  & $623.9$  &$31.74$ & $0.889$ & $28.26$ & $0.772$ & $27.40$ & $0.728$ & $25.50$ & $0.763$ \\      
		LapSRN   & $4$ & $813$  & $149.4$  &$31.54$ & $0.885$ & $28.19$ & $0.772$ & $27.32$ & $0.728$ & $25.21$ & $0.756$ \\
		\midrule[0.5pt]
		CARN     & $4$ & $1592^{*}$ & $65.4$   &$32.13$ & $0.894$ & $28.60$ & $0.781$ & $27.58$ & $0.735$ & $26.07$ & $0.784$ \\  
		DRCN     & $4$ & $1774$ & $9788.7$ &$31.53$ & $0.885$ & $28.02$ & $0.767$ & $27.23$ & $0.723$ & $25.14$ & $0.751$ \\ 
		$s$-LWSR$_{64}$(Ours)   & $4$ & $2277$ & $131.1$   &$32.28$ & $0.896$ & $28.34$ & $0.780$ & $27.61$ & $0.738$ & $26.19$ & $0.791$ \\
		$s$-LWSR$_{64}+$(Ours)  & $4$ & $2277$ & $131.1$   &$32.42$ & $0.898$ & $28.42$ & $0.782$ & $27.69$ & $0.739$ & $26.39$ & $0.795$ \\
		D-DBPN   & $4$ & $10426$& $590.2$  &$32.47$ & $0.898$ & $28.82$ & $0.786$ & $27.72$ & $0.740$ & $26.38$ & $0.795$ \\ 
		EDSR     & $4$ & $43090$& $2482.0$ &$32.46$ & $0.897$ & $28.80$ & $0.788$ & $27.71$ & $0.742$ & $26.64$ & $0.803$ \\
		\bottomrule[2pt]
	\end{tabular}
\end{table*}

{\bfseries Activation Removal.}
In addition to the compression block, we further remove several activation layers to retain more details in the very model with small size. Note that the thinner channel design of the small model limits its learning ability. How to retain more accurate information of input becomes a crucial factor regarding to better performance. Hence, we imply the strategy of removing some activation layers to keep more information. 

\begin{figure}[!htbp]
	\centering
	\subfigure[]{
		\includegraphics[width=4.1cm]{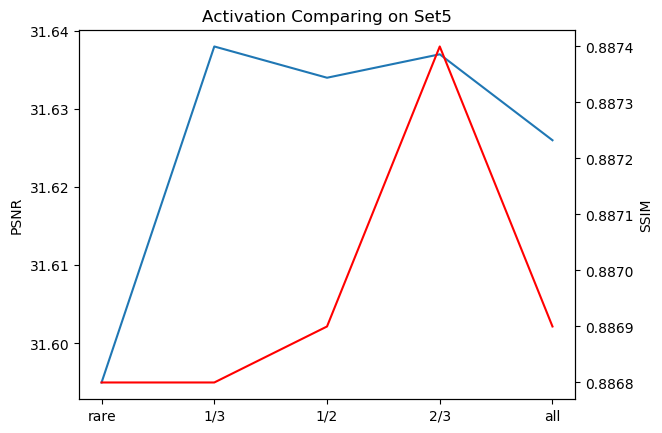} 
	}
	\subfigure[]{
		\includegraphics[width=4.1cm]{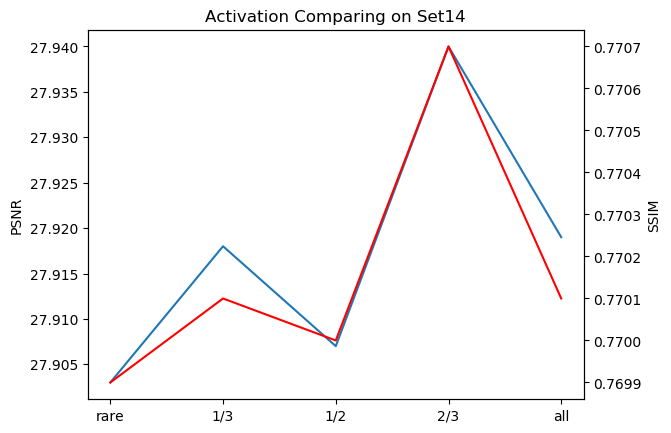}
	}
	\subfigure[]{
		\includegraphics[width=4.1cm]{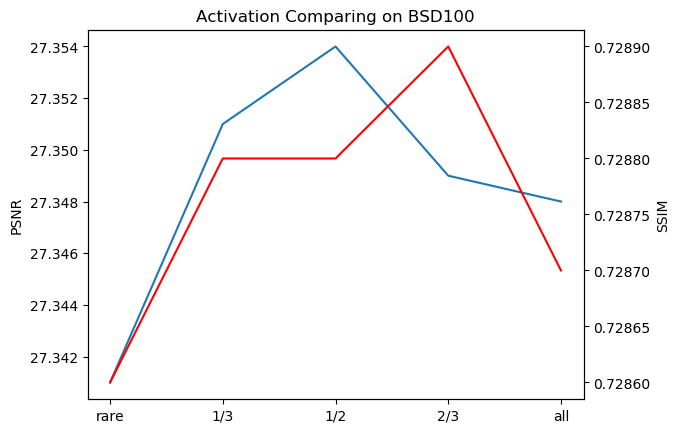} 
	}
	\subfigure[]{
		\includegraphics[width=4.1cm]{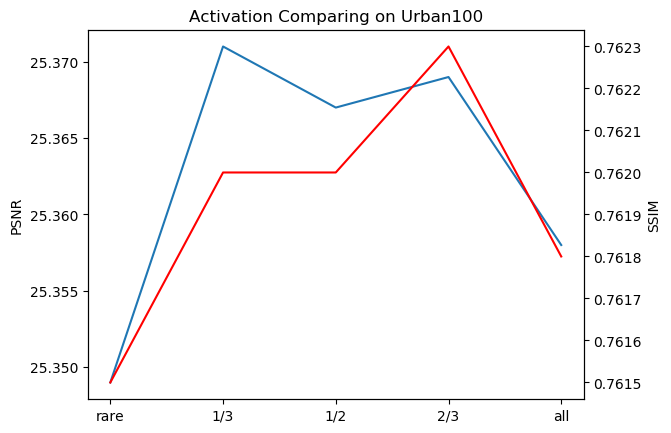} 
	}
	\caption{The PSNR and the SSIM values of s-LWSR$_{16}$ with different ratios of activation layers removal on $Set5$ \cite{bevilacqua2012low}, $Set14$ \cite{zeyde2010single}, $BSD100$ \cite{martin2001database}, and $Urban100$ \cite{huang2015single}. There are five settings: rare(only conv layers keep activation), 1/3 activation, half activation, 2/3 activation, and all activation. In details, the comparing results indicate that s-LWSR$_{16}$ with part activation layers ablation possesses significant advantage in both PSNR and SSIM.} 
	\label{activation}
\end{figure}

To evaluate our opinion, some activation layers are discarded from the model-s-LWSR$_{16}$. More comparing experiments are done for the purpose that how the model change with the reduce of activation layers. Actually, we decrease activation operations with the setting: rare (only first and last convolutional layers) kept, 1/3 kept, 1/2 kept, 2/3 kept and all. It can be inferred from the results that the removal of the moderate number of activate layers brings the beneficial effect on the small SR model. Even with a few activation layers, our model can still achieve comparable results. What's more, with the increasing of parameters, on the contrary, the removing operation results in a worse performance. We can see from the chart that better performance is achieved in all middle setting(like 1/3, 2/1, or 2/3) compared with rare and all activation layers kept. The final outputs are illustrated in Fig. \ref{activation}. Our final s-LWSR model imply half activation setting to obtain a better balance between PSNR and SSIM.

\subsection{Ablation Study}\label{as}
In s-LWSR, is the newly introduced information pool really works for final performance? To answer this question, we design the ablation experiments. Besides, the chosen channels are evaluated by different setting to better evaluate effects.

For the purpose of acquiring  multi-dimensional information of inputs, chosen layers of the front half model are concatenated as the information pool which provides hybrid features to latter layers. From the perspective of information utilization, the more details are involved, the better performance of model achieves. However, over recurrence leads to overfitting. We respectively compare the performance of different setting. Moreover, s-LWSR with $16$, $32$ and $64$ channels are all involved for clarifying the effect of model size. 

As shown in Table \ref{tab3}, the existence of information pool slightly increases SSIM score and PSNR. We mark the best scores in red color. The benefit exists among all three settings and performs better with the increase of channel number. This trend is related to learning ability and more parameters. Because of minor filtered operations, former layers extract more accurate and useful information from input. As a result, chosen layers bring these better details into the information pool and are transferred to the latter layers. Because there are skip connections without information pool, the improvements are limited in a rather small level.

We further make contrast experiments in s-LWSR$_{16}$ to check out the effect of layers involved in the information pool. Note that skip connections play equal influence as the information pool, we just compare three extreme conditions of front half: all involved, half, and none. In Table \ref{tab3}, all SR results are shown in \ref{tab3}. From the table, we can inform that s-LWSR$_{16}$ obtains better generations in mostly datasets where the only exception is marked in blue color. Even though, SR generations achieve equal PSNR score, the SSIM provides additional evidence of the effect. We attribute the advantage of s-LWSR$_{16}$ to reasonable using of the information. To be specific,  s-LWSR$_{16}$ without pool transfers information by the skip-residual mechanism which transmits given layer to fixed ones. However, our pool block gathers layers from various channels, which concatenates multi-dimensions information. Referring to  s-LWSR$_{16}$ with all former layers, repetitive features of adjacent layers lead to overfitting.

\begin{table*}[!htbp]
	\begin{center}
		\small
		\caption{The comparison of the original s-LWSR and two derivatives transformed on the depth or the width. The changes of parameters and PSRN are illustrated.}
		\label{tab3}
	    \begin{tabular}{c|c|cc|cc|cc|cc}
	    	\toprule[2pt]
	    	&& \multicolumn{2}{c}{Set5} & \multicolumn{2}{c}{Set14} & \multicolumn{2}{c}{B100} & \multicolumn{2}{c}{Urban100}\\
	    	Algorithm &	Scale & PSNR & SSIM & PSNR & SSIM & PSNR & SSIM & PSNR & SSIM \\
	    	\midrule[1.5pt]
	    	$s$-LWSR$_{16}$(normal)  & $4$ & $31.63$ & $0.8869$ & $27.92$ & $0.7701$ & $27.35$ & ${\color{red}0.7287}$ & $25.36$ & $0.7618$ \\
	    	$s$-LWSR$_{16}$(no-pool)  & $4$ & $31.63$ & $0.8868$ & $27.92$ & $0.7696$ & $27.35$ & $0.7284$ & $25.36$ & $0.7616$ \\	
	    	$s$-LWSR$_{16}$(pool- former 11 layers)  & $4$ & $31.63$ & ${\color{blue}0.8871}$ & $27.90$ & $0.7698$ & $27.34$ & $0.7286$ & $25.36$ & $0.7616$ \\	    	
	    	\hline
	    	$s$-LWSR$_{32}$(normal)  & $4$ & ${\color{red}32.02}$ & ${\color{red}0.893}$ & ${\color{red}28.15}$ & $0.776$ & ${\color{red}27.52}$ & $0.734$ & ${\color{red}25.87}$ & $0.779$ \\		
	    	$s$-LWSR$_{32}$(no-pool)  & $4$ & $31.97$ & $0.892$ & $28.12$ & $0.776$ & $27.51$ & $0.734$ & $25.86$ & $0.779$ \\		
	    	\hline
	    	$s$-LWSR$_{64}$(normal)  & $4$ & $32.23$ & $0.896$ & ${\color{red}28.34}$ & $0.780$ & $27.61$ & $0.738$ & ${\color{red}26.19}$ & ${\color{red}0.791}$ \\		
	    	$s$-LWSR$_{64}$(no-pool)  & $4$ & $32.23$ & $0.896$ & $28.32$ & $0.780$ & $27.61$ & $0.738$ & $26.13$ & $0.790$ \\		
		    \bottomrule[2pt]
		\end{tabular}
	\end{center}
\end{table*}

\begin{figure*}
	\centering
	\includegraphics[scale=0.6]{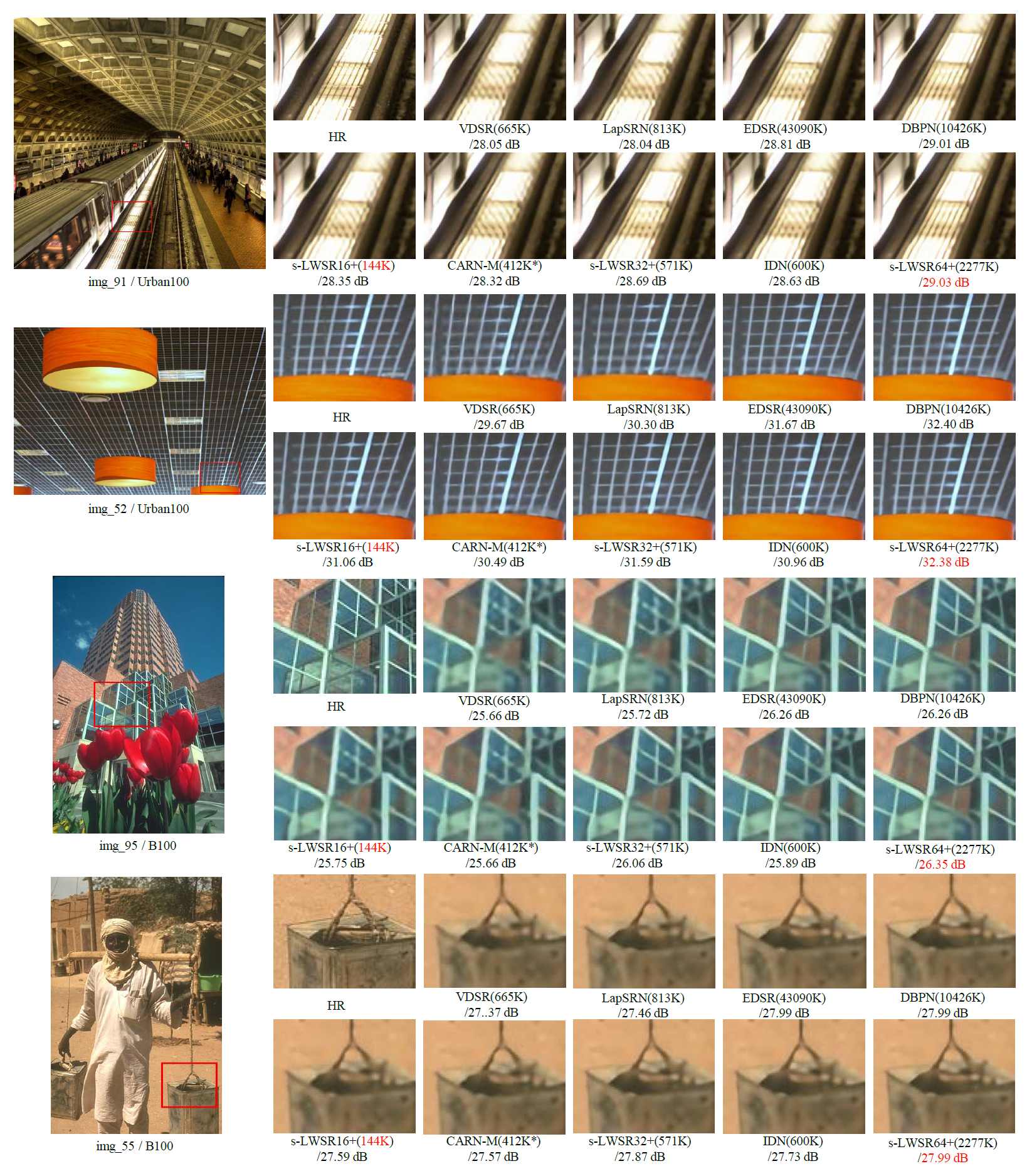}
	\caption{Qualitative comparison with leading algorithms: VDSR, LapSRN, EDSR, DBPN, IDN, and CARN on $4\times$ task. From the figure, we can point out that s-LWSR achieves outstanding performance when there is similar parameters. With the adding of more channels, s-LWSR show persistent increasing in learning ability. As shown, s-LWSR64 supass most of existing SR model in PSNR and SSIM on condition of less parameters.}
	\label{cont2}
\end{figure*}

\subsection{Comparison with State-of-the-art Models}
To confirm the learning ability of our proposed network, we compare our model with several state-of-the-art methods: SRCNN \cite{Chao2014Learning}, FSRCNN \cite{10.1007/978-3-319-46475-6_25}, CARN\cite{article2}, VDSR \cite{Kim2016Accurate}, MemNet \cite{tai2017memnet}, IDN \cite{hui2018fast}, LapSRN \cite{Lai2017Deep}, DRCN \cite{kim2016deeply}, DBPN \cite{Haris_2018_CVPR}, and EDSR \cite{Lim2017Enhanced}. We conduct the evaluation experiments through two frequently-used image quality metrics: the PSNR and the SSIM. Most pre-trained models are directly based on the $DIV2K$. Here, it is noting that that the DBPN and the CARN are trained with extra images as they declaring in their papers. Accordingly, test datasets are $Set5$, $Set14$, $B100$, and $Urban100$. In this paper, all methods are only performed for the $4\times$ SR task.

For precise comparison, we separate these algorithms into three sections based on their sizes: $0-500K$, $500K-1000K$, and $1000K+$. It is worth noticing that the CARN actually contains three times parameters in the main network than that asserted in the single scale-up model. Here, we only compare with the asserted size. {color{red}To maximize the performance of SR generations, we adopt the self-ensemble strategy which is widerly used in EDSR, RCAN. Moreover, to separate enhanced version with original SR, the $+$ is added behind initial name.} In the first section, s-LWSR$_{16}+$ performs a little worse than CARN-M, while greatly surpasses SRCNN and FSRCNN. However, the total number of parameters and operations in s-LWSR$_{16}+$ is only half of the asserted value of the CARN-M. In the second section, s-LWSR$_{32}+$ outperforms all the competitors. It can be concluded from Table \ref{tab1} that s-LWSR$_{32}+$ demonstrates great advantages on both model size and accuracy in a large margin. Referring to the last section, s-LWSR$_{64}+$ performs similarly with the DBPN and the EDSR. Meanwhile, the size of our model is distinctly smaller on both parameters and operations. Besides, the outputs of $4\times$ enlargement are visually exhibited in Fig. \ref{cont2}. In general, the comparison suggests that our model has a strong capability in the SR generation, weather in the lightweight model size or better accuracy.

\section{Conclusion}
In this paper, we propose a super lightweight SR network: s-LWSR. To facilitate the implementation on mobile devices, we compress our model to only $144K$ parameters while the s-LWSR achieves a satisfying performance. Base on the symmetric architecture, we propose an information pool with skip-connection mechanism to comprehensively incorporate the multi-level information. Besides, we further explore s-LWSR with more channels and remove certain ratios of activation layers to achieve comparable performance with leading SR models. In addition, we introduce a compression module to further reduce the model size to the ideal scale. The extensive experiments demonstrate that our model performs better than other state-of-the-art lightweight SR algorithms, with a relatively smaller model size.

% if have a single appendix:
%\appendix[Proof of the Zonklar Equations]
% or
%\appendix  % for no appendix heading
% do not use \section anymore after \appendix, only \section*
% is possibly needed

% use appendices with more than one appendix
% then use \section to start each appendix
% you must declare a \section before using any
% \subsection or using \label (\appendices by itself
% starts a section numbered zero.)
%

%\appendices
%\section{Proof of the First Zonklar Equation}
%Appendix one text goes here.

% you can choose not to have a title for an appendix
% if you want by leaving the argument blank
%\section{}
%Appendix two text goes here.

% use section* for acknowledgment
\section*{Acknowledgment}
Thanks for the helpful suggestions from reviewers of the ACM MM'19. This research was funded by the National Natural Science Foundation of China No. 91546201, No. 71801232, No. 71501175, and No. 61702099, the Open Project of Key Laboratory of Big Data Mining and Knowledge Management, Chinese Academy of Sciences, the Fundamental Research Funds for the Central Universities in UIBE (No.CXTD10-05), and the Foundation for Disciplinary Development of SITM in UIBE.

% Can use something like this to put references on a page
% by themselves when using endfloat and the captionsoff option.
\ifCLASSOPTIONcaptionsoff
  \newpage
\fi

% trigger a \newpage just before the given reference
% number - used to balance the columns on the last page
% adjust value as needed - may need to be readjusted if
% the document is modified later
%\IEEEtriggeratref{8}
% The "triggered" command can be changed if desired:
%\IEEEtriggercmd{\enlargethispage{-5in}}
%\newpage
% references section
\bibliographystyle{IEEEtran}
\bibliography{s-LWSR}
\end{document}